\documentclass[aps,prb,twocolumn,showpacs,showkeys]{revtex4}
\usepackage{graphicx}
\usepackage{epsfig}
\usepackage{amssymb}
\usepackage{bm}

\def\be{\begin{equation}} \def\ee{\end{equation}} \def\bea{\begin{eqnarray}}
\def\eea{\end{eqnarray}} \def\nnb{\nonumber}

\begin{document}
\setcounter{page}{1}
\title[]{$pp\to pp\pi^0$ near threshold in pionless 
effective field theory}
\author{Shung-ichi \surname{Ando}}
\email{sando@color.skku.ac.kr}
\affiliation{Department of Physics, Sungkyunkwan University, 
Suwon 440-746, Korea}

\begin{abstract}
In this talk, we review our recent calculation 
for the $pp\to pp\pi^0$ reaction
near threshold in pionless effective field theory
with a di-baryon and external pions.
\end{abstract}



\maketitle

\section{INTRODUCTION}

The study of neutral pion production in proton-proton collision
near threshold, $pp\to pp\pi^0$,
has been inspired by precise measurements of the near-threshold
cross section~\cite{metal-prl90,metal-npa92}.
Surprisingly, the measured cross section
turned out to be $\sim$ 5 times larger
than the early theoretical predictions~\cite{kr-pr66}.
Subsequently, some mechanisms to account for
the threshold experimental data
have been suggested in model calculations~\cite{lr-prl93}.

Heavy-baryon chiral perturbation theory (HB$\chi$PT)
is a low-energy effective field theory (EFT) of QCD
and provides us
a systematic perturbation scheme
in terms of $Q/\Lambda_\chi$ where
$Q$ denotes small external momentum and/or symmetry breaking term
$m_\pi$ and $\Lambda_\chi$ denotes the chiral scale
$\Lambda_\chi=4\pi f_\pi\simeq 1$ GeV:
$f_\pi$ is the pion decay constant.
Though many works on the $pp\to pp\pi^0$ reaction near
threshold in HB$\chi$PT have been 
done~\cite{petal-prc96,cetal-prc96,setal-prc97,detal-plb99,aetal-plb01,ketal-07,vkmr-plb96,bkm-epja99}
(for a recent review, see Ref.~\cite{h-pr04} and references therein),
some issues in theoretically describing the process
have not been fully clarified.
In the next-to-leading order (NLO)
HB$\chi$PT calculations~\cite{petal-prc96,cetal-prc96,setal-prc97},
a significant enhancement of the off-shell $\pi\pi NN$
vertex function obtained from the NLO HB$\chi$PT Lagrangian
is found. However,
the two-body (one-pion-exchange) matrix element
with the off-shell $\pi\pi NN$ vertex
is almost exactly canceled with the one-body
matrix element.
Thus the experimental data cannot be reproduced in the NLO calculations.
In the next-to-next-to leading order (NNLO) 
HB$\chi$PT calculations~\cite{detal-plb99,aetal-plb01,ketal-07},
a significant contribution
comes out of the NNLO corrections
and a moderate agreement with the experimental data
is obtained~\cite{aetal-plb01}.
However, the chiral series
based on the standard Weinberg's counting rules~\cite{weinberg}
shows poor convergence.

A modification of the original Weinberg's counting rules
to account for the large momentum transfer,
$k\simeq \sqrt{m_\pi m_N}$
where $m_\pi$ and $m_N$ are the pion and nucleon masses, respectively,
is discussed in Ref.~\cite{cetal-prc96}.
The production operators at NLO using the modified counting rules
are estimated, and it was reported that
the NLO contributions exactly cancel
among themselves~\cite{hk-prc02}.
Recently, some detailed issues
for the loop calculations, such as
a concept of reducibility~\cite{letal-epja06},
a representation invariance of the chiral fields
among the loop diagrams,
and a proper choice of the heavy-nucleon
propagator~\cite{hw-plb07},
were also studied.

Meanwhile, it is known that the energy dependence of the
experimental data can be well reproduced in terms of
the final state interaction and the phase space~\cite{metal-npa92}.
A ``minimal'' formalism to
take account of these two features
would be a {\it pionless} theory,
in which
virtual pions exchanged between the two nucleons
are integrated out;
in this pionless theory, the one-pion exchange,
two-pion exchange and contact terms in HB$\chi$PT
are subsumed in a contact term.
Furthermore, after taking these two features into account,
the difference between the theory and experiment
appears in the overall factor and
the experimental data can be easily reproduced
by fitting an unknown constant
that appears in a contact vertex.
In this work~\cite{sa-epja07} we employ a pionless EFT
with a di-baryon~\cite{bs-npa01,ah-prc05} \footnote{
We have studied $np\to d\gamma$ cross section at BBN 
energies\cite{achh-prc06}
and neutron-neutron fusion process\cite{ak-plb06}
employing this formalism. }
and external pions~\cite{bs-npa03}
to calculate the total cross section of
the $pp\to pp\pi^0$ process.

\section{Pionless effective Lagrangian}

An effective Lagrangian without virtual pions
and with a di-baryon and external pions
for describing the $pp\to pp\pi^0$ reaction may read
\bea
{\cal L} =
  {\cal L}_N
+ {\cal L}_s
+ {\cal L}_{Ns\pi}
+ {\cal L}_{NN}^P \, ,
\eea
where 
${\cal L}_N$ is the standard one-nucleon Lagrangian in heavy-baryon
formalism  where the ``external'' pions are nonlinearly realized.
${\cal L}_{s}$ is 
for the $^1S_0$ channel di-baryon field,
${\cal L}_{Ns\pi}$ 
represents the contact interaction of an external pion,
a di-baryon and two nucleons,
and ${\cal L}_{NN}^P$ is for the two-nucleon 
$^3P_0$ channel. 
The effective Lagrangian for the two-nucleon part
may read~\cite{bs-npa01,ah-prc05,crs-npa99,fms-npa00}
\bea
{\cal L}_s &=&
\sigma_s s_a^\dagger\left[iv\cdot D
+\frac{1}{4m_N}[(v\cdot D)^2-D^2]
+\delta_s\right] s_a
\nnb \\ &&
-y_s\left[s_a^\dagger (N^TP^{(^1S_0)}_aN) + \mbox{h.c.}\right],
\\
{\cal L}_{Ns\pi} &=&
\frac{\tilde{d}_\pi^{(2)}}{\sqrt{8 m_Nr_0}}
\left\{
i\epsilon_{abc}
s_a^\dagger
\left[
N^T\sigma_2 \vec{\sigma}\cdot i
(\stackrel{\to}{D}
-\stackrel{\leftarrow}{D})
\tau_2\tau_b N
\right] 
\right.
\nnb \\ && \left.
\times (iv\cdot \Delta_c)
+ \mbox{\rm h.c.}
\right\}
\, ,
\\
{\cal L}_{NN}^P &=&
C_2^0 \delta_{ij}\delta_{kl}\frac14
\left(N^T{\cal O}_{ij,a}^{1,P}N\right)^\dagger
\left(N^T{\cal O}_{kl,a}^{1,P}N\right)
+ \cdots \, ,
\eea
with
${\cal O}_{ij,a}^{1,P} =
i(\stackrel{\leftarrow}{D}_i P_{j,a}^{(P)}
-P_{j,a}^{(P)} \stackrel{\rightarrow}{D}_i)$ and
$P_{i,a}^{(P)} = \frac{1}{\sqrt{8}}\sigma_2 \sigma_i \tau_2\tau_a$. 
$s_a$ is the $^1S_0$ channel di-baryon field and
$\sigma_{s}$ is the sign factor $\sigma_{s}=\pm 1$
which we fix below.
$v^\mu$ is a velocity vector 
$v^\mu=(1,\vec{0})$ and
$D_\mu$ is the covariant derivative.
$\delta_s$ is the mass difference
between the di-baryon mass $m_s$
and two-nucleon mass, 
$m_{s} = 2 m_N + \delta_{s}$.
$y_{s}$ is the coupling constant
of the di-baryon and two-nucleon interaction.
$P_i^{({}^1S_0)}= \frac{1}{\sqrt8}\tau_2\tau_a\sigma_2$.
$d_\pi^{(2)}$ is an unknown low energy constant (LEC) of the
(external) pion-(spin singlet) dibaryon-nucleon-nucleon ($\pi sNN$)
interaction.
$r_0$ is the effective range in the $^1S_0$ ($pp$) channel,
and $\Delta^\mu=\frac{\tau_a}{2}\Delta^\mu_a$.
$C_2^0$ is the LEC for the 
$NN$ scattering in the $^3P_0$
channel.

\begin{figure}
\begin{center}
\includegraphics[width=9cm]{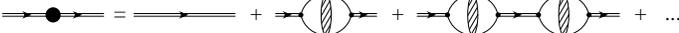}
\caption{
Diagrams for the dressed di-baryon propagator
including the Coulomb interaction.
\label{fig;dibaryon-propagator}
}
\end{center}
\end{figure}

Now we calculate the $S$- and $P$-wave $NN$
scattering amplitudes to fix the LECs in the two-nucleon part.
In Fig.~\ref{fig;dibaryon-propagator},
diagrams for the dressed $^1S_0$ channel di-baryon propagator
are shown where the two-nucleon bubble diagrams
including the Coulomb interaction
are summed up to the infinite order.
\begin{figure}[t]
\begin{center}
\includegraphics[width=3.7cm]{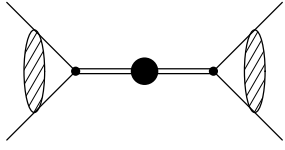}
\caption{
Diagram for the $S$-wave $pp$ scattering amplitude
with the Coulomb interaction.
\label{fig;NNamplitudes}}
\end{center}
\end{figure}
In Fig.~\ref{fig;NNamplitudes},
a diagram of the $S$-wave $pp$ scattering
amplitude with the Coulomb interaction is shown
and
thus we have the $S$-wave scattering amplitude as
\bea
\lefteqn{i{\cal A}_{s} =
(-iy_{s}\psi_0)\left[iD_{s}(p)\right](-iy_{s}\psi_0)}
\nnb \\ &&
=
i \frac{4\pi}{m_N}\frac{C_\eta^2e^{2i\sigma_0}}{
-\frac{4\pi\sigma_{s}\delta_{s}^R}{m_Ny_{s}^2}
-\frac{4\pi\sigma_{s} p^2}{m_N^2y_{s}^2}
-\alpha m_N h(\eta)
-ip\, C_\eta^2} \, ,
\eea
where 
$\psi_0 = C_\eta e^{i\sigma_0}$ and  
$\sigma_0$ is the $S$-wave Coulomb phase shift
$\sigma_0={\rm arg}\,\Gamma(1+i\eta)$.
$D_s(p)$ is the dressed di-baryon propagator and
$\delta_s^R$ is the renormalized mass difference.  
$h(\eta) = Re\, \psi(i\eta)-{\rm ln}\eta$, 
$Re\, \psi(\eta) =
\eta^2 \sum_{\nu=1}^\infty \frac{1}{\nu(\nu^2+\eta^2)}
-\gamma$, 
$\gamma= 0.5772\cdots$, and
\bea
C_\eta^2 = \frac{2\pi \eta}{e^{2\pi\eta}-1}\, ,
\ \ \
\eta= \frac{\alpha m_N}{2p}\, .
\eea
The $S$-wave amplitude ${\cal A}_s$ is
given in terms of the effective range parameters
as
\bea
i{\cal A}_s =
i \frac{4\pi}{m_N}\frac{C_\eta^2e^{2i\sigma_0}}{
-\frac{1}{a_C} + \frac12 r_0 p^2 + \cdots
-\alpha m_N h(\eta)
-ip\, C_\eta^2} \, ,
\eea
where $a_C$ is the scattering length,
$r_0$ is the effective range,
and the ellipsis represents
the higher order corrections.
Now it is easy to match the LECs
with the effective range parameters.
Thus we have $\sigma_{s} = -1$ and
\bea
y_s &=& \pm \frac{2}{m_N}\sqrt{\frac{2\pi}{r_0}}\, ,
\label{eq;ys}
\\ 
D_s(p) &=& \frac{m_Nr_0}{2}\frac{1}{\frac{1}{a_C}
-\frac12r_0 p^2 +\alpha m_N h(\eta) +ip\, C_\eta^2}\, .
\eea
We note that the sign of the LEC $y_s$ cannot be determined
by the effective range parameters.

\begin{figure}[t]
\begin{center}
\includegraphics[width=8.0cm]{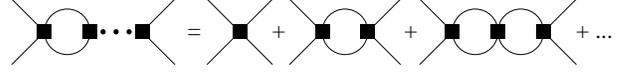}
\caption{
Diagrams for the $P$-wave $NN$ scattering.
\label{fig;pwavescattering}}
\end{center}
\end{figure}

In Fig.~\ref{fig;pwavescattering},
diagrams for the $P$-wave $NN$ scattering
are shown.
Because the momenta of the two protons are quite large
for the pion production reaction,
the two-proton bubble diagrams
are summed up to the infinite order.
The scattering amplitude for
the ${}^3P_0$ channel is obtained as
\bea
i{\cal A}_p &=& \frac{4\pi}{m_N}\frac{ip^2}
 {\frac{4\pi}{m_NC_2^0}-ip^3} \, .
\eea
The LEC $C_2^0$ is
fixed by the phase shift of the ${}^3P_0$ channel at
pion production threshold,
$\delta_p(p_{th})\simeq -7.5^\circ$ at
$p_{th}\simeq \sqrt{m_\pi m_N}$.
Thus we have
\bea
\frac{4\pi}{m_NC_2^0} \simeq p_{th}^3\cot\delta_p(p_{th}) \, .
\eea

\section{
Amplitudes for $pp\to pp\pi^0$ near threshold }

\begin{figure}[t]
\begin{center}
\includegraphics[width=6.0cm]{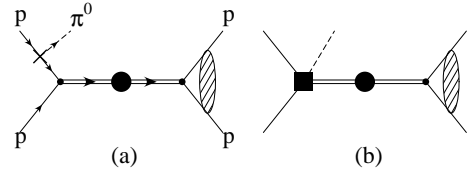}
\caption{
Diagrams for $pp\to pp\pi^0$  near threshold
with the strong and Coulomb final state interactions
and without the initial state interaction.
\label{fig;diagrams-non-FSI}}
\end{center}
\end{figure}
\begin{figure}[t]
\begin{center}
\includegraphics[width=8.0cm]{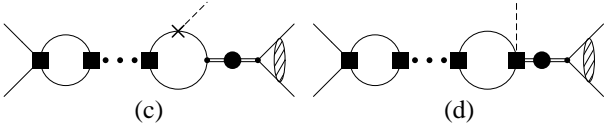}
\caption{
Diagrams for $pp\to pp\pi^0$
with the strong initial and the strong and Coulomb final state interactions.
\label{fig;diagrams-FSI}}
\end{center}
\end{figure}
In Figs.~\ref{fig;diagrams-non-FSI}
and \ref{fig;diagrams-FSI},
we show diagrams for $pp\to pp\pi^0$
near threshold.
In diagram (a) in Fig.~\ref{fig;diagrams-non-FSI}
and (c) in Fig.~\ref{fig;diagrams-FSI},
the pion is emitted from the one-body $\pi NN$ vertex.
In the diagram (b) in Fig.~\ref{fig;diagrams-non-FSI}
and (d) in Fig.~\ref{fig;diagrams-FSI},
the pion is emitted from the $\pi sNN$ contact vertex 
which is proportional to the LEC $\tilde{d}_\pi^{(2)}$.
The one-body amplitude from the (a) and (c) diagrams
and the two-body (contact) amplitude from the (b) and (d) diagrams
are obtained as
\bea
\lefteqn{i{\cal A}_{(a+c)} = 
-\frac{4\pi g_A}{m_N^2f_\pi}
\frac{1}{1-\frac{m_NC_2^0}{4\pi}ip^3}}
\nnb \\ && \times
\frac{C_{\eta'} e^{i\sigma_0}p}
{\frac{1}{a_C}-\frac12r_0p'^2
+\alpha m_N h(\eta') +ip'\, C_{\eta'}^2} \,,
\label{eq;Aac}
\\
\lefteqn{i{\cal A}_{(b+d)} =
4\sqrt{\frac{2\pi}{m_N}}\frac{\tilde{d}_\pi^{(2)}}{f_\pi}
\frac{1}{1-\frac{m_NC_2^0}{4\pi}ip^3}
} 
\nnb \\ && \times
\frac{C_{\eta'}e^{i\sigma_0}\omega_q p}
{\frac{1}{a_C} -\frac12 r_0 p'^2
+\alpha m_N h(\eta') +ip'\, C_{\eta'}^2}
\label{eq;Abd}
\eea
where 
$2\vec{p}$ and $2\vec{p}'$ are the relative three
momenta between incoming and outgoing two protons, respectively;
$p=|\vec{p}|$ and $p'=|\vec{p}'|$.
$\eta'=\alpha m_N/(2p')$ and
$\omega_q$ is the energy of outgoing pion,
$\omega_q=\sqrt{\vec{q}^2+m_\pi^2}$:
$\vec{q}$ is the outgoing pion momentum.
We note that there remain no unknown parameters
in the amplitudes except for
the LEC $\tilde{d}_\pi^{(2)}$
in the two-body (contact) amplitude in Eq.~(\ref{eq;Abd}).

\begin{figure}[t]
\begin{center}
\includegraphics[width=3.7cm]{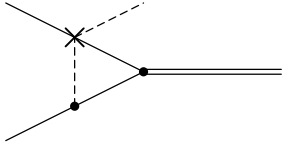}
\caption{
Diagram for a one-pion exchange contribution
to the $pp\to pp\pi^0$ process for estimating
the LEC $\tilde{d}_\pi^{(2)}$ in the contact vertex.
\label{fig;diagrams-loops}}
\end{center}
\end{figure}

Now we estimate an order of magnitude of the LEC $\tilde{d}_\pi^{(2)}$
from HB$\chi$PT.
We here consider a one-pion-exchange (OPE)
diagram shown in Fig.~\ref{fig;diagrams-loops}.
This diagram is the lowest order OPE contribution
in the standard Weinberg counting rules.
We include a higher order (relativistic) correction
to the $\pi\pi NN$ vertex which is found to
be important~\cite{bkm-epja99} and is,
in the modified counting rules,
of the same order
as the lowest order diagram.

The effective chiral Lagrangian 
to calculate the isoscalar $\pi\pi NN$ interaction
in the diagram in Fig.~\ref{fig;diagrams-loops},
reads~\cite{fmms-ap00}
${\cal L}_{\pi N} =  
{\cal L}_{\pi N}^{(2)}
+ {\cal L}_{\pi N}^{(3)}
+ \cdots$ where 
\bea
{\cal L}_{\pi N}^{(2)} &=& N^\dagger
\left[
c_1 {\rm Tr}(\chi_+)
+ \left(
\frac{g_A^2}{2m_N}-4c_2\right)
(v\cdot \Delta)^2
\right. \nnb \\ && \left. \frac{}{}
-4c_3\Delta\cdot \Delta
\right] N +\cdots\, ,
\label{eq;L2}
\eea
and
${\cal L}^{(3)}_{\pi N}$ is the relativistic correction 
to the term proportional to $(v\cdot\Delta)^2$ in 
Eq.~(\ref{eq;L2}). 
The values of the LECs $c_1$, $c_2$ and $c_3$
are fixed in the tree-level calculations~\cite{bkm-npb95} 
as
\bea
c_1 = -0.64\, ,
\ \ \
c_2 = 1.79 \, ,
\ \ \
c_3 = -3.90\
\ [\mbox{\rm GeV$^{-1}$}]\, .
\label{eq;c123}
\eea

The value of the LEC $\tilde{d}_\pi^{(2)}$
from the loop diagram
in Fig.~\ref{fig;diagrams-loops}
is obtained as
\bea
\tilde{d}_\pi^{(2)} &\simeq& \pm \frac{\sqrt{2\pi}g_A}{32m_\pi^{3/2}}
\frac{m_\pi^2}{f_\pi^2}\left(
-4c_1
+2c_2
-\frac{3g_A^2}{16m_N}
+c_3
\right)
\nnb \\ 
&\simeq& \pm 0.140\ \ \mbox{\rm fm$^{5/2}$}\, ,
\label{eq;dpi2}
\eea
where the different signs for
$\tilde{d}_\pi^{(2)}$
have been obtained because of the LEC $y_s$
in Eq.~(\ref{eq;ys}).

\section{Numerical results and Summary}

The total cross section of $pp\to pp\pi^0$ near threshold
is calculated using the formula
\bea
\sigma = \frac12 \int^{q^{max}}_0 \!\!\!\! dq \frac{d\sigma}{dq},
\ \ 
\frac{d\sigma}{dq} = \frac{1}{v_{lab}}
\frac{m_Nq^2p'}{16(2\pi)^3\omega_q}\sum_{spin}|{\cal A}|^2,
\label{eq;sigma}
\eea
with
$p' =|\vec{p}'| \simeq \sqrt{m_N(T-\sqrt{m_\pi^2+q^2})-q^2/4}$,
and  
$q^{max} \simeq \sqrt{\frac{T^2-m_\pi^2}{1+\frac{T}{2m_N}}}$. 
$T$ is the initial total energy $T\simeq\vec{p}^2/m_N$ and
$v_{lab}\simeq 2p/m_N$.
We have expanded the proton energies
in the phase factor
in terms of $1/m_N$ and kept
up to the $1/m_N$ order.
${\cal A}$ is the amplitude
${\cal A} = {\cal A}_{(a+c)} + {\cal A}_{(b+d)}$
where ${\cal A}_{(a+c)}$ and ${\cal A}_{(b+d)}$
are obtained
in Eqs.~(\ref{eq;Aac}) and (\ref{eq;Abd}), respectively.

\begin{figure}[t]
\begin{center}
\includegraphics[width=8.0cm]{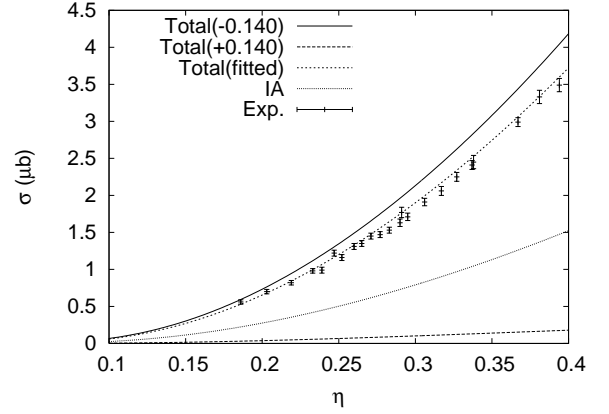}
\caption{
Estimated total cross section of $pp\to pp\pi^0$
as a function of $\eta_\pi=|\vec{q}|_{max}/m_\pi$.
See the text for details.
\label{fig;CS}}
\end{center}
\end{figure}

In Fig.~\ref{fig;CS} we plot our results
for the total cross section
as a function of
$\eta_\pi= q^{max}/m_\pi$.
The solid curve and long-dashed curve have been obtained
by using $\tilde{d}_\pi^{(2)}=\pm 0.140$ fm$^{5/2}$
fixed from the one-pion exchange diagram
in Fig.~\ref{fig;diagrams-loops}
in the previous section.
The LEC $\tilde{d}_\pi^{(2)}$ is also fixed by using
the experimental data as
\bea
\tilde{d}_\pi^{(2)fitted} = - 0.12 \, , \ \ +0.55 \ \mbox{\rm fm}^{5/2}\, ,
\eea
where we have two values of $\tilde{d}_\pi^{(2)}$ with different signs.
The short-dashed curve
is obtained by
using $\tilde{d}_\pi^{(2)fitted}= -0.12$ fm$^{5/2}$.
The dotted line corresponds to the case where
only the contribution from the one-body
amplitude ${\cal A}_{(a+c)}$ is considered.
The experimental data are also included
in the figure.

We find that
the experimental data are
reproduced reasonably well
with the value of $\tilde{d}_\pi^{(2)} = -0.14$ fm$^{5/2}$.
By contrast,  we obtain almost vanishing total cross sections
with the value $\tilde{d}_\pi^{(2)}=+0.140$ fm$^{5/2}$
because the two-body amplitude
with $\tilde{d}_\pi^{(2)}=+0.140$ fm$^{5/2}$
is almost canceled with the amplitude from the one-body
contribution.
On the other hand,
for the whole energy range the experimental
near-threshold cross section data
are well reproduced with the use of  the fitted parameter
$\tilde{d}_\pi^{(2)fitted}=-0.12$ fm$^{5/2}$.
We also find that
approximately a half of the
observed cross section comes from
the one-body (IA) amplitude
in the pionless theory.

In this work we calculated the total cross section
for $pp\to pp\pi^0$ near threshold
in pionless EFT with the di-baryon and external pion fields.
The leading one-body amplitude and subleading contact
amplitude were obtained including the strong initial
state interaction and the strong and Coulomb final-state
interactions.
After we fix the LECs for the $NN$ scatterings,
there remains only one unknown constant, $\tilde{d}_\pi^{(2)}$,
in the amplitude.
We estimated it from the one-pion exchange
diagram in the pionful theory.
Although this method does not allow us
to fix the sign of $\tilde{d}_\pi^{(2)}$,
we have found that one of the two choices for $\tilde{d}_\pi^{(2)}$
leads to the cross sections that agree with
the experimental data reasonably well.
On the other hand, the whole range
of the experimental data near threshold
can be reproduced by adjusting the only
unknown LEC in the theory, $\tilde{d}_\pi^{(2)}$.
As discussed in Introduction, this is an expected result
because the energy dependence of the experimental total cross section
is known to be well described by the phase factor
and the final-state interaction~\cite{metal-npa92},
which have been taken into account in this work,
and the overall strength of the cross section can be
adjusted by the value of $\tilde{d}_\pi^{(2)}$.
This feature would be the same
in the NNLO HB$\chi$PT calculations
because an unknown constant
appears in the contact $\pi NNNN$ vertex
and can be adjusted
so as to reproduce the experimental data
though there are many other corrections coming out
of the pion loop diagrams.

\begin{acknowledgments}
The author would like to thank 
the organizers for the conference APPC10, 
August 21-24, 2007, Pohang, Korea for hospitality.
This work is supported by Korean Research Foundation and The
Korean Federation of Science and Technology Societies Grant
founded by Korean Government
(MOEHRD, Basic Research Promotion Fund): the Brain Pool program
(052-1-6) and KRF-2006-311-C00271.

\end{acknowledgments}

\end{document}